\documentclass[12pt,pra,aps,amssymb,amsfonts,amsmath,tightenlines]{revtex4}%aps instead of rmp
\usepackage[caption = false]{subfig}
\usepackage{dsfont}
\usepackage{graphicx}
\usepackage{amssymb,amsfonts,amsthm}
\usepackage{color}
\usepackage{verbatim}
\usepackage[normalem]{ulem}
\usepackage{enumerate}
\usepackage{mathrsfs}
\usepackage{bm,float}

\usepackage{textgreek}
\usepackage{lgreek}

\usepackage{nicefrac}

\newcommand{\half}{\mbox{$\textstyle \frac{1}{2}$}}

\newcommand{\re}{\mbox{$\rm e$}}

\newcommand{\rd}{\mbox{$\rm d$}}

\begin{document}

\title{Quantum formalism for cognitive psychology}
\author{Dorje~C.~Brody}

\affiliation{
School of Mathematics and Physics, University of Surrey, 
Guildford GU2 7XH, United Kingdom
}

\date{\today}
%ABSTRACT HERE
\begin{abstract}
\noindent 
The cognitive state of mind concerning a range of choices to be made can effectively be modelled in terms of an element of a high-dimensional Hilbert space. The dynamics of the state of mind resulting form information acquisition is characterised by the von Neumann-L\"uders projection postulate of quantum theory. This is shown to give rise to an uncertainty-minimising dynamical behaviour equivalent to the Bayesian updating, hence providing an alternative approach to characterising the dynamics of cognitive state that is consistent with the free energy principle in brain science. The quantum formalism however goes beyond the range of applicability of classical reasoning in explaining cognitive behaviours, thus opens up new and intriguing possibilities. 
\vspace{-0.2cm}
\\
%KEY WORDS HERE
%\keywords{Key Words}
%\subclass{MSC code\and JEL classification code}
\end{abstract}

\maketitle

%%%%%%%%%%%%%%
\section*{Introduction}
\label{sec:1} 

%\begin{quotation}
%\noindent 
%Such sentiments are seldom thoroughly stifled unless by reference
%to the doctrine of chance, or, as it is technically termed, the
%Calculus of Probabilities. Now this Calculus is, in its essence,
%purely mathematical; and thus we have the anomaly of the most
%rigidly exact in science applied to the shadows and spirituality
%of the most intangible in speculation. \\ \hspace*{\fill}
%{\scriptsize E. A. Poe, ``The Mystery of Marie Rog\^et''}
%\end{quotation}

%\noindent \textit{Introduction}. 
The present paper is concerned with the use of Hilbert space techniques, so 
successfully implemented in characterising behaviours and properties of 
quantum systems \cite{Isham}, to model cognitive `psychology' in the sense 
to be defined 
below. The paper, on the other, is \textit{not} concerned with whether quantum 
effects, such as interference, violation of the Bell inequality, or the use of 
complex numbers, might play a role in psychology or in brain activities. There 
are differing opinions on the matter 
\cite{Penrose,Vitiello,Tegmark,Plotnitsky,Ozawa}, 
but these will not be addressed here, for, the use of Hilbert space techniques 
in modelling cognitive behaviour, in themselves, do not necessarily require 
functioning of brain neurophysiology to be quantum mechanical. 
What does concern us will include the tensor product structure of the Hilbert 
space, entanglements, the superposition principle, the projection postulate, 
and decoherence. While some may view 
these to be intrinsically quantum-mechanical effects, I will show that they 
are in fact intrinsic to any probabilistic system modelled on a Hilbert space 
--- an idea that dates back to the pioneering work of Rao \cite{Rao}. At 
any rate, my purpose here is to illustrate how it is both natural and effective to 
implement Hilbert space techniques in modelling cognitive human behaviours. 
Such a proposal, in itself, is not new (see, e.g., \cite{Busemeyer,Busemeyer2} 
and references cited therein). My main contribution is to introduce a formalism 
that allows for a systematic treatment of the \textit{dynamics} of cognitive 
state of mind, from which predictions can be made. 

The key idea to be explored is that the state of mind of a person, to be defined 
more precisely, can efficiently be represented in terms of a vector in a 
high-dimensional Hilbert space, which in turn is a tensor product of 
lower-dimensional Hilbert spaces. However, before 
turning to technical discussion, I would like to illustrate the significance of the 
superposition principle in this context 
through simple examples. The first concerns the tossing of a 
fair coin. When a coin is tossed, but the outcome not yet revealed, no one will dispute 
the state of the coin, which is a macroscopic classical object: it is \textit{either} in 
the `head' state \textit{or} in the `tale' state. In fact, even before the coin is tossed, it 
will be known to 
all that the state of the coin \textit{will be} either head or tail. However, a 
person who is in the position to guess the outcome is having a different state of mind. 
Until the moment when a choice is made, the person's mind is in the state of 
\textit{neither} head \textit{nor} tail -- it is in a state of a 
superposition. Thus there is a dissonance between the 
objective reality of the state of the coin and the state of the mind attempting to 
determine the state of the coin. 

Another example is seen in children's hand game of rock, scissors and paper. 
Again, until the last moment when the decision is made as to which hand to 
choose, the person's mind is in the state of a superposition between the three 
alternatives. Of course, with any decision, a person may come with a predetermined 
outcome --- as in Groucho Marx's whatever it is, I'm against it!  But until when 
that decision 
is made, the fact remains that the mind is in an indefinite state. Only at the moment 
when the decision is made, for example to declare that the outcome of the coin 
tossing is head, or choosing the scissors hand, the state `collapses' into one of the 
definite states. This situation is very much analogous to the measurement 
process in quantum theory. 

With these heuristics in mind, the present paper will be organised as follows. 
I begin by defining what I mean by the state of mind, which essentially will be 
a Hilbert-space representation of the probability assignments to the totality of 
choices available to the person. A given set of choices will then be modelled 
as an observable acting on this Hilbert space. The idea I put 
forward is closely related to the proposal made by Busemeyer and his 
collaborators \cite{Busemeyer0}, and that by Khrennikov and his 
collaborators \cite{Khrennikov}. One difference to the latter, however, is that I 
model cognitive behaviours directly using the quantum formalism, rather than 
starting from the quantum theory and then deduce implications in behavioural 
modelling by reduction. As a consequence, I am able to gain certain insights 
into human behaviour that go beyond what is commonly understood in 
cognitive psychology. I examine how the state of mind changes 
when a person acquires noisy information that is relevant to decision 
making. My approach will be consistent with the cybernetics framework of 
Wiener as an attempt to understand the dynamics of living systems 
\cite{Wiener}. I will use the von Neumann-L\"uders projection postulate in 
quantum mechanics to arrive at the change in the state of mind, and show 
that the result agrees with the classical formulation using the Bayes formula; 
in agreement with related previous findings \cite{Khrennikov}. 
Further, I show that the projection postulate gives rise to an evolution 
that on average minimises future `surprises', and hence is consistent with 
the free-energy principle widely adopted in brain science \cite{Friston0}. I 
then consider what happens when noisy information arrives in continuous 
time. When the dynamical evolution of the state of mind resulting from a 
sequential von Neumann-L\"uders projection postulate is reduced to a 
projective Hilbert space, the dynamical equation reveals an important 
feature, namely, that each state of zero uncertainty, having vanishing 
Shannon-Wiener entropy, acts as an 
attractor to the dynamics. The result provides a geometric explanation of 
certain characteristics of human behaviour seen in psychology literature 
(such as confirmation bias), which I have previously called the tenacious 
Bayesian behaviour \cite{Brody2}. The work presented here demonstrates 
the fact that if a person's opinion is strongly skewed towards one of the 
false alternatives, and if all choice observables are commutative, then 
even if partial information about the truth is revealed and the person 
behaves rationally in accordance with Bayesian updating, it is very 
difficult for the person to escape from the initial misconception. In other 
words, adherences to false narratives commonly observed in today's 
society is not due to irrational behaviours. Following the proposal in 
\cite{Busemeyer}, I then consider what happens if choice observables 
are not commutative. This is motivated by the observation that many 
empirical observations concerning cognitive human behaviours cannot 
be fully described by use of a commuting set of observables. The 
existence of incompatible observables can be used to rescue a person 
fallen into a false attractor; a possibility that is unavailable by use of 
purely classical reasonings.

\section*{Decision making and the state of mind} 

%\vspace{0.2cm} 
%\noindent \textit{Decision making and the state of mind}. 
Throughout the paper I will be concerned with the cognitive process of 
decision making, which will be a topic distinct from what is known as 
statistical decision theory, for which there are excellent treatise 
\cite{DeGroot,Berger}. A decision making occurs when a person is unsure 
from which alternative to choose; but I will be using the term `decision making' 
in a broad sense to include a person's uncertain point of view on a topic, for 
which there are multiple views, and for which there may not be a need to 
choose one particular alternative. At any rate, this uncertainty, which is largely 
due to lack of sufficient information, can be modelled in the form of a set of 
probabilities that represents the likelihoods of different alternatives being 
selected. Suppose that a decision needs to be made to choose one out of 
$N$ alternatives. (The number of alternatives can be infinite, or even 
uncountable --- the formalism extends naturally to these cases, but for 
simplicity I will consider the finite case here.) 
At a given moment in time, let $p_k$ denote the likelihood 
that the $k$th alternative is selected. If $p_k=1$ for a value of $k$  then the 
mind is in a definite state in relation to this decision. Hence for a given 
decision, the set of numbers $(p_1,p_2,\ldots,p_N)$ represents the state of 
mind in relation to that decision making. 

Of course, at any moment in time one faces a multitude of decisions, not just one, 
some of which are intertwined with each other while others are independent. In 
probability theory, such a situation is modelled by means of a joint probability for 
the totality of decisions. Alternatively, though equivalent, the situation can be 
modelled on a Hilbert space by use of the square-root map: $p_k \to \psi_k = 
\sqrt{p_k}$. Clearly, the vector with components $\{\psi_k\}$, in the basis 
$|e_k\rangle = (0,0,0,\ldots,1,0,\ldots,0)$ with only the $k$th element nonzero, 
is an element of an $N$-dimensional real Hilbert space ${\cal H}^N$. Thus, in 
terms of the Dirac notation the state can be expressed in the form of a 
superposition $|\psi\rangle = \sum_k \psi_k |e_k\rangle$. If there is a second 
decision to be made out of $M$ alternatives, then the state of a person's mind 
in relation to these two choices is represented by an element of the tensor 
product ${\cal H}^N\otimes{\cal H}^M$. This tensor product structure arises 
solely from statistical dependencies of two decisions, when modelled on a 
Hilbert space. 

This construction extends for an arbitrary number of decisions to be made. 
With this in mind, I define the state of mind of a person facing a range of 
alternatives to consider, at any moment in time, to be an element of the 
tensor product ${\cal H}=\otimes_{l=1}^K{\cal H}_l$, where $K$ is the number 
of distinct decisions. If two decisions can be made independently, then the 
component of the state vector belonging to the corresponding subspace of 
${\cal H}$ will be in a product state. Otherwise, a state is entangled. As a simple 
example, consider a pair of binary decisions, for example, whether to take 
fish or meat for the main course, and whether to take red or white wine to 
accompany 
the dinner. Writing $|F\rangle$ and $|M\rangle$ for the food choices, 
and similarly $|R\rangle$ and $|W\rangle$ for the wine selections, if the state of 
mind of a person is $|\psi\rangle = c_1 |FW\rangle + c_2 |MR\rangle$, then the 
person will choose fish with white wine with probability $c_1^2$, and meat with 
red wine with probability $c_2^2=1-c_1^2$; but no other option will be chosen. 
This is evidently an entangled state, which collapses to one or the other 
alternatives at the moment (or before) the waiter arrives and takes the order. 

In this Hilbert space formulation, a given choice can be modelled by a real 
symmetric matrix, whose dimension is the number of 
alternatives. Such a matrix corresponds to observables in quantum mechanics. 
I will assume, for now, that all such `observables' or `choices' are compatible 
in the sense that the matrix representations can be 
diagonalised simultaneously. What this means is that at any given time, an 
arbitrary number of decisions can be made simultaneously. It is then evident 
that no state of mind, whether entangled or not, can violate laws of classical 
probability, and hence no state can violate, in particular, Bell's inequalities. 
Later in the paper, however, I will consider the case where choice 
observables are incompatible. 

The eigenvalues of choice observables then label different alternatives. This 
is analogous to quantum observables when it concerns labelling outcomes of a 
single measurement. However, observables in quantum theory have a second 
role apart from representing measurement outcomes: they generate dynamics. 
As a consequence, the differences of observable eigenvalues have direct 
physical consequences, and hence they cannot be chosen arbitrarily. It appears, 
in contrast, that the differences of eigenvalues of the choice observables have 
no significance: the results of a selection, such as choosing a hand in 
the game of rock, scissors and paper, can be labelled by means of any three 
distinct numerical values, merely as place keepers so that statistical analysis 
can be applied. It will be shown below, however, that when 
it concerns the dynamics of the state of mind, the eigenvalue differences do 
play an important role, and hence they cannot be chosen arbitrarily, just as in 
quantum theory. 

\section*{Dynamics}

%\vspace{0.2cm} 
%\noindent \textit{Dynamics}. 
Having established the framework for representing the cognitive state of 
mind, it will be of interest to explore how the state changes in time. To this 
end I will be working under the hypothesis that a given state $|\psi\rangle$ 
of a person's mind changes only by transfer of information. It is, of course, 
possible that an isometric motion analogous to unitary motion of quantum 
theory that does not exchange information can change the state, and if so 
this will be given by an orthogonal transformation. However, without a clear 
physical or psychological evidence indicating the existence of such a symmetry, 
I will not consider this possibility, and focus instead on universally acknowledged 
empirical fact that information acquisition (or loss) changes states of minds. 
The question is, in which way? 

To understand dynamics, I will be borrowing ideas from communication theory. 
Focusing on a single decision to start with, let ${\hat X}$ denote the decision or 
choice observable, with eigenvalues $\{x_k\}_{k=1,\ldots,N}$. These eigenvalues 
for now merely label different alternatives. The eigenstate $|x_j\rangle$ of 
${\hat X}$, satisfying ${\hat X}|x_j\rangle =x_j|x_j\rangle$, thus represents 
the state of mind in which the $j$th alternative has been chosen. Prior to an 
alternative being chosen, the state is in a superposition $|\psi\rangle = 
\sum_k c_k |x_k\rangle$. The state will change when the person acquires 
information relevant to decision making. This information is rarely perfect. In 
communication theory, anything that obscures finding the value of the quantity 
of interest is modelled in terms of noise. Let ${\hat\varepsilon}$ denote this noise. 
Here, ${\hat\varepsilon}$ can take discrete values, or more commonly continuous 
values. I will consider the latter case so that ${\hat\varepsilon}$ acts on an infinite 
dimensional Hilbert space ${\cal H}^\infty$ distinct from the state space 
${\cal H}^N$. The noise arises from external environments ${\cal E}$. For 
simplicity I will assume that the state of noise is pure, and is given by 
$|\eta\rangle =\eta(y) \in {\cal H}^\infty$, although a mixed state can equally be 
treated. Then initially the state of mind of a person attempting to make a decision 
and the state of the noise-inducing environment is disentangled, and together is 
given by the product state $|\psi\rangle|\eta\rangle$. 

Acquisition of partial information relevant to decision making can 
then be modelled by observing the value of 
\[
{\hat\xi} = {\hat X} + {\hat\varepsilon} . 
\] 
Here, the sum is taken in the tensor-product space ${\cal H}^N\otimes
{\cal H}^\infty$. To understand the sum, consider the case in which 
${\hat\varepsilon}$ 
is finite and can take three values $\varepsilon_1,\varepsilon_2,\varepsilon_3$, 
while the decision is binary, represented by the values $x_0$ and $x_1$. Then 
${\hat\xi}$ is a $6\times6$ matrix with the eigenvalues $x_0+\varepsilon_1$, 
$x_0+\varepsilon_2$, $x_0+\varepsilon_3$, $x_1+\varepsilon_0$, $x_1+ 
\varepsilon_1$, $x_1+\varepsilon_3$. In general, the eigenvalues of ${\hat\xi}$ 
are highly (typically $N$-fold) degenerate. The form that ${\hat\xi}$ takes is of 
course nothing more than a signal-plus-noise decomposition in classical 
communication theory \cite{Wiener}. The 
`signal' term, more generally, will be a function $F({\hat X})$ of ${\hat X}$, but for 
simplicity I will assume the function to be linear because the choice of $F(x)$ is 
context dependant. 

Once the value of the information-providing observable ${\hat\xi}$ is measured, 
the initially-disentangled state 
\[ 
|\psi\rangle|\eta\rangle = \sum_k \sqrt{p_k} \, \eta(y) |x_k\rangle 
\] 
becomes an entangled state. In quantum mechanics, the transformation of the 
state after measurement is given by the von Neumann-L\"uders projection 
postulate. That is, writing 
\[
{\hat\Pi}_\xi = \sum_k \delta(y-\xi+x_k) |x_k\rangle \langle x_k| 
\] 
for the projection operator onto the subspace of ${\cal H}^N\otimes
{\cal H}^\infty$ spanned by the eigenstates of ${\hat\xi}$ with the eigenvalue 
$\xi$, the projection postulate asserts that the state of the system after 
information acquisition is 
\[ 
{\hat\rho}_\xi = \frac{{\hat\Pi}_\xi|\psi\rangle|\eta\rangle
\langle\psi|\langle\eta|{\hat\Pi}_\xi} {{\rm tr}\left({\hat\Pi}_\xi|\psi\rangle 
|\eta\rangle\langle\psi|\langle\eta|{\hat\Pi}_\xi\right)} . 
\]
A short calculation then shows that this is given more explicitly by 
\[
{\hat\rho}_\xi = \frac{\sum_{k,l} \sqrt{p_k p_l} \, \eta(\xi-x_k) \eta(\xi-x_l) \, 
|x_k\rangle \langle x_l|}{\sum_m p_m \eta^2(\xi-x_m)} . 
\]
Two interesting observations that follow are in order. First, the density 
matrix by construction is a projection operator onto a random pure state 
$|\psi(\xi)\rangle$ given by 
\[
|\psi(\xi)\rangle = \sum_k \sqrt{\pi_k(\xi)}\, |x_k\rangle , 
\] 
where $\xi$ is the random variable with the density $p(y)=\sum_m p_m 
\eta^2(y-x_m)$ modelled by the observable ${\hat\xi}$ along with the initial state 
$|\psi\rangle|\eta\rangle$. That is, given the state $|\psi\rangle|\eta\rangle$, the 
probability of the measurement outcome of the observation lying in the 
interval $[y,y+\rd y]$ is given by $p(y)\rd y$. Second, the coefficients of the 
random pure state agrees with the conditional probability of the choice given 
by the Bayes formula: 
\[
\pi_k(\xi) = \frac{p_k\,\eta^2(\xi-x_k)}{\sum_m p_m \eta^2(\xi-x_m)} . 
\] 
That is, $\pi_k(\xi)$ is the probability that the $k$th alternative is chosen, 
conditional on observing the value $\xi$ of ${\hat\xi}$. It follows that the von 
Neumann-L\"uders projection postulate of quantum theory not only gives 
the correct classical result (as already observed in \cite{Ozawa,Khrennikov} 
with a different construction) 
but also provides a simple geometric interpretation of the 
Bayes formula. This follows because the L\"uders state $|\psi(\xi)\rangle$ 
associated to a degenerate measurement outcome $\xi$ is given by the 
orthogonal projection of the initial state onto Hilbert subspace associated to 
this outcome. Hence $|\psi(\xi)\rangle$ is the closest state on 
the constrained subspace in terms of the Bhattachayya distance \cite{BH} 
to the initial state $|\psi\rangle|\eta\rangle$. 

It is worth remarking that an alternative interpretation of the von 
Neumann-L\"uders projection postulate can be given in terms of the so-called 
free energy principle \cite{KF1}. Intuitively, this principle asserts that the 
change in the state of mind follows a path that on average minimises elements 
of surprises. In the present context, the degree of surprise can be measured in 
terms of the level of uncertainty. Suppose that the state of mind after information 
acquisition becomes ${\hat\mu}_\xi$ that is different from the L\"uders state 
${\hat\rho}_\xi$. Then the level of uncertainty associated with the choice 
observable ${\hat X}$ resulting from ${\hat\mu}_\xi$, 
conditional on the observed value $\xi$ of ${\hat\xi}$, is given by 
\[ 
{\rm tr}\left( \left[{\hat X}-{\rm tr}({\hat X}{\hat\mu}_\xi)\right]^2{\hat\rho}_\xi \right) 
= {\rm tr}\left( {\hat X}^2{\hat\rho}_\xi \right) - \left( {\rm tr}\left( {\hat X}{\hat\rho}_\xi 
\right) \right)^2 + \left( {\rm tr}\left( {\hat X}{\hat\delta}_\xi\right) \right)^2,
\]
where I have written ${\hat\delta}_\xi={\hat\mu}_\xi-{\hat\rho}_\xi$ for the deviation. 
Because the first two terms on the right side is the conditional variance of ${\hat X}$, 
which is positive and is independent of ${\hat\mu}_\xi$, to minimise the expected 
uncertainty, and hence the surprise, for all 
${\hat X}$ and $\xi$, it has to be that ${\hat\delta}_\xi=0$. It follows that among all 
the states consistent with the observation, L\"uders state is unique in that it 
minimises the expected level of future surprise, as measured by the uncertainty. 

I might add parenthetically that a psychologist wishing to predict the statistics 
of the behaviour of a person who has acquired information relevant to decision 
making will \textit{a priori} not know the observed value of $\xi$. Hence in this 
case the density matrix ${\hat\rho}_\xi$ has to be averaged over $\xi$, 
but the denominator of ${\hat\rho}_\xi$ is just the density $p(y)$ for $\xi$, so the 
averaged density matrix is given by 
\[ 
{\mathbb E}[{\hat\rho}_\xi] = 
\sum_{k,l} \sqrt{p_k p_l} \, \Lambda(\omega_{kl}) \, |x_k\rangle \langle x_l| , 
\] 
where $\omega_{kl}=x_k-x_l$ and 
\[ 
\Lambda(\omega) = \int_{-\infty}^\infty \eta(y) \eta(y-\omega) \rd y  . 
\] 
Evidently, $0\leq\Lambda(\omega_{kl})\leq1$ 
and $\Lambda(\omega_{kk})=1$ for all 
$k,l$, but because the initial state of mind $|\psi\rangle\langle\psi|$ in this basis 
has the matrix elements $\{\!\sqrt{p_kp_l}\}$, it follows that an external observer 
(e.g., a psychologist) will perceive a decoherence effect whereby the off-diagonal 
elements of the reduced density matrix are damped. 

It is at this point that I wish to comment on the numerical values of the differences 
$\{\omega_{kl}\}$. While there is no reason why $\Lambda(\omega)$ should be 
monotonic in $\omega$ (unless $\eta(y)$ is unimodal), it will certainly be 
the case that the decoherence effect is more pronounced for large values of 
$\omega$. That is, $\Lambda(\omega)\ll1$ for $\omega\gg1$. For the same 
token, the values of $\omega_{kl}$ will directly affect the conditional probabilities 
$\{\pi_k(\xi)\}$. Therefore, while the values of $\omega_{kl}$, and hence those of 
$x_k$, can be chosen arbitrarily to describe the statistics of the initial state of 
mind, once dynamics is taken into account (what happens \textit{after} information 
acquisition), it becomes evident that they cannot be chosen arbitrarily. 

A better intuition behind this observation can be gained by reverting back to 
ideas of signal detection in communication theory. For this purpose, consider 
a binary decision. Supposed that the two eigenvalues of ${\hat X}$, labelling 
the two decisions, are chosen to be, say, $\pm0.1$ and suppose that the 
noise distribution $\eta^2(y)$ is normal centred at zero, with a small standard 
deviation. In this case, observed outcomes of ${\hat\xi}$ will most likely take 
values close to zero. As a consequence, a single observation of ${\hat\xi}$ 
will reduce on average the initial uncertainty only by a very small amount. In 
contrast, suppose that the two eigenvalues of ${\hat X}$ are chosen to be 
$\pm10$, 
but the noise is the same as before. Then the observation will almost certainly 
yield the outcome that is close to $+10$ or $-10$. Hence the uncertainty in 
this case has been reduced to virtually zero after a single observation. This 
extreme example shows how it is not possible to label different choice 
alternatives by arbitrary numerical numbers, while at the same time to 
adequately model the dynamics of the state of mind. 

In the event where a model $\eta(y)$ for the state of noisy environment 
exists, it is possible in principle to estimate the eigenvalue differences 
$\{\omega_{ij}\}$ by studying how much a person's views shifted from the 
acquisition of the noisy information. This is because the average reduction 
of uncertainty, as measured by entropy change or the decoherence rate, 
is determined by the eigenvalue differences $\{\omega_{ij}\}$. 

\section*{Sequential updating}

%\vspace{0.2cm} 
%\noindent \textit{Sequential updating}. 
I have illustrated how the cognitive state of mind of a person in relation to a 
given choice changes after a single acquisition of information. A more interesting, 
as well as realistic, situation concerns a sequential updating of the state 
of mind as more and more noisy information is revealed. In this case the 
information-providing observable ${\hat\xi}_t$ is a time series. As a simple 
example that naturally extends the previous one I will consider the following 
time series 
\[ 
{\hat\xi}_t = {\hat X}t+{\hat\varepsilon}_t , 
\] 
where I will assume that the noise term ${\hat\varepsilon}_t$ is modelled by 
a standard Brownian motion $\{B_t\}$ 
multiplied by the identity operator of the Hilbert space ${\cal H}^\infty$. The 
`signal' component, more generally, can be given by $\int_0^t F_s({\hat X}) 
\rd s$, but again for simplicity I will assume that the function $F_t(x)$ is 
linear for all $t$. In fact, even more generally, the range of alternatives 
${\hat X}$ itself can also be time dependent, but I will not consider this case 
here. 

In this example, what happens to the state of mind can be worked out by 
discretising the time and taking the limit. Starting from time zero, over a 
small time increment $\rd t$ the initial state $|\psi\rangle|\eta\rangle$ is 
projected to the L\"uders state ${\hat\Pi}_{\xi_{{\rm d}t}}|\psi\rangle|\eta\rangle$,
suitably normalised, in accordance with the projection postulate. 
In this case, the noise is normally distributed with mean zero and variance 
$\rd t$, so that $\eta(y)$ is just the square-root of the corresponding 
Gaussian density function. Then after another time interval $\rd t$ we apply 
the projection operator again, and repeat the procedure till time $t$. Finally, 
taking the limit, a calculation shows that the L\"uders state, after monitoring 
the observable ${\hat\xi}_t$ up to time $t$, is given by 
\[ 
|\psi(\xi_t)\rangle = \frac{1}{\sqrt{\Phi_t}} 
\sum_k \sqrt{p_k} \, \re^{ \frac{1}{2}x_k \xi_t - \frac{1}{4} 
x_k^2 t} |x_k\rangle , 
\] 
where $\xi_t=Xt+B_t$ and $X$ is the random variable represented on the 
Hilbert space ${\cal H}^N$ by the operator ${\hat X}$ along with the initial 
state $|\psi\rangle$, and $\Phi_t = \sum_k p_k \re^{ \frac{1}{2}k_k 
\xi_t - \frac{1}{4} x_k^2 t}$ gives the normalisation. 

Because we have an explicit expression that monitors the change in the 
state of mind as information is revealed, there is \textit{a priori} no reason to 
identify the differential equation to which $|\psi(\xi_t)\rangle$ is the solution. 
Nevertheless, the exercise of working out the dynamical equation provides 
several new insights worth discussing. The detailed mathematical steps 
required here to work out the dynamics has been outlined in \cite{BH2}, so 
I shall not repeat this. It suffices to say that the L\"uders state is a function 
of $t$ and $\xi_t$, where the latter is a Brownian motion with a random 
drift. Hence the relevant calculus to apply is that of Ito: one Taylor expands 
$|\psi(\xi_t)\rangle$ in $t$ and $\xi_t$, and retain leading-order terms, bearing 
in mind that $(\rd\xi_t)^2=\rd t$. Then it follows that 
\[ 
\rd|\psi(\xi_t)\rangle = -\frac{1}{8}({\hat X}-\langle{\hat X}\rangle_t)^2 
|\psi(\xi_t)\rangle \, \rd t + \frac{1}{2} ({\hat X}-\langle{\hat X}\rangle_t) 
|\psi(\xi_t)\rangle \, \rd W_t , 
\] 
where $\langle{\hat X}\rangle_t=\langle\psi(\xi_t)|{\hat X}|\psi(\xi_t)\rangle/
\langle\psi(\xi_t)|\psi(\xi_t)\rangle$ and where 
\[ 
\rd W_t = \rd \xi_t - \langle{\hat X}\rangle_t \, \rd t . 
\] 
The process $\{W_t\}$ defined in this way is in fact a standard Brownian 
motion, known as the innovations process \cite{Kailath}. This process 
has the interpretation of revealing new information. That is, while 
the time series $\{{\hat\xi}_t\}$ contains new as well as previously known 
information about the impending choice to be made, the process $\{W_t\}$ 
merely contains information that was not known previously. 

There are two important observations that follow. First, the evolution of 
the state of mind is not directly generated by the noise $\{B_t\}$, nor by the 
observation $\{\hat\xi_t\}$. Rather, it is the innovations process that drives 
the dynamics. But this is the case only if the state of mind changes in such 
a way to continuously minimise uncertainties. Because the expectation of 
the cumulative uncertainty is the entropy \cite{BH2}, it follows that according 
to the present framework, the tendency towards low entropy states 
required in biology \cite{KF1}, which forms the basis of the free energy 
principle, emerges naturally. In particular, the implication here based on the 
projection postulate is that the state of mind changes only in accordance 
with the arrival of new information; it will not change spontaneously of its 
own otherwise. Second, while the analysis 
presented here can be deduced as a result of standard least-square estimation 
theory \cite{Kailath,Wonham}, I have derived these results using the von 
Neumann-L\"uders projection postulate of quantum theory. It follows that 
the informationally efficient dynamical behaviour, in the sense of minimising 
surprises, of a system, is applicable not only to the state of mind but also to 
quantum systems. An analogous point of view, based on the free energy 
principle, has recently been proposed elsewhere \cite{KF2}. 

I might add that 
for the purpose of psychological modelling, the averaged reduced density 
matrix ${\hat\rho}_t={\mathbb E}[{\hat\rho}_{\xi_t}]$ can be seen to obey 
the dynamical equation 
\[ 
\frac{\partial{\hat\rho}_t}{\partial t} = {\hat X}{\hat\rho}_t{\hat X} - 
\frac{1}{2}\left( {\hat X}^2{\hat\rho}_t+{\hat\rho}_t{\hat X}^2 \right). 
\] 
This, of course, is just the Lindblad equation generated by the decision 
${\hat X}$. 

\section*{Projecting down the dynamics} 

%\vspace{0.2cm} 
%\noindent \textit{Projecting down the dynamics}. 
One advantage of working with the mathematical formalism of quantum 
theory in modelling psychological states of minds is the deeper insights 
that it can uncover (cf. \cite{Busemeyer}). 
To this end I note that although I have defined the state 
of mind as a vector in Hilbert space, what I have in mind really is a 
projective Hilbert space consisting of rays through the origin of the Hilbert 
space. The idea is as follows. In probability, one can say, for instance, that 
the likelihood of an event happening is 0.3, or three out of ten, or 30\% --- 
all of these statements convey the same idea. The total probability being 
equal to one is merely a 
convenient convention that does not carry any significance. Putting it 
differently, working with the convention that the expectation of any 
decision ${\hat X}$ in a state $|\psi\rangle$ is given by the ratio $\langle 
{\hat X}\rangle = \langle\psi|{\hat X}|\psi\rangle/\langle\psi|\psi\rangle$, it 
is evident that the expectation values are independent of the overall 
scaling of the state $|\psi\rangle$ by a nonzero constant. Hence the Hilbert 
space vector $|\psi\rangle$ carries one psychologically irrelevant degree 
of freedom. When this degree of freedom is eliminated by the identification 
$|\psi\rangle \sim \lambda|\psi\rangle$ for any $\lambda\neq0$, one 
arrives at a projective Hilbert space, otherwise known as the real projective 
space. This is a real manifold ${\mathfrak M}$ 
of dimension $N-1$, endowed with a 
Riemannian metric induced by the underlying probabilistic rules given by 
the von Neumann-L\"uders projection postulate \cite{BH4}. 

\begin{figure}[t]

  \centering
       \subfloat[Negative gradient flow of the variance]
       {\includegraphics[width=0.48\textwidth]{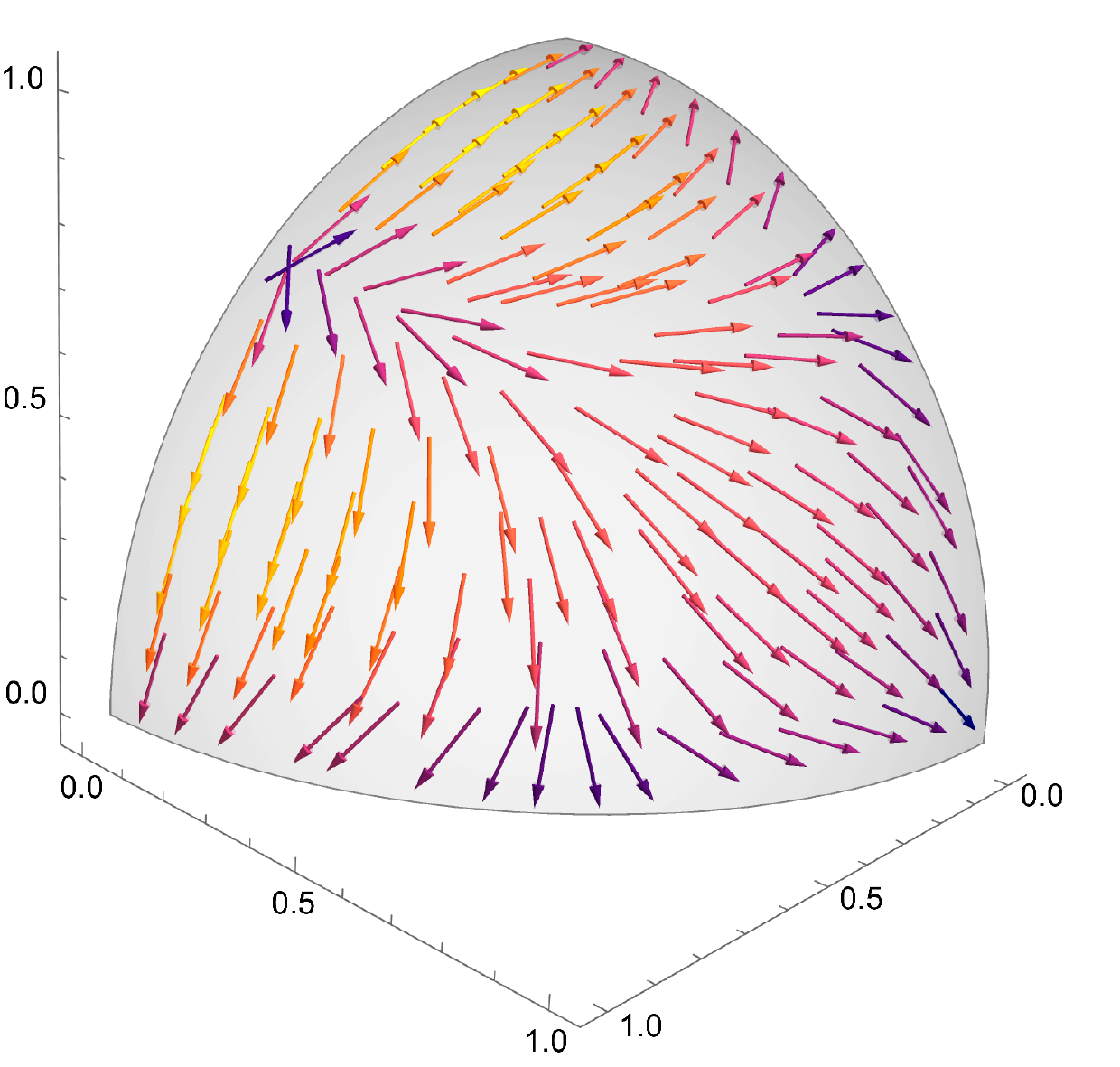}}\hfill
       \subfloat[Gradient flow of the mean]
       {\includegraphics[width=0.48\textwidth]{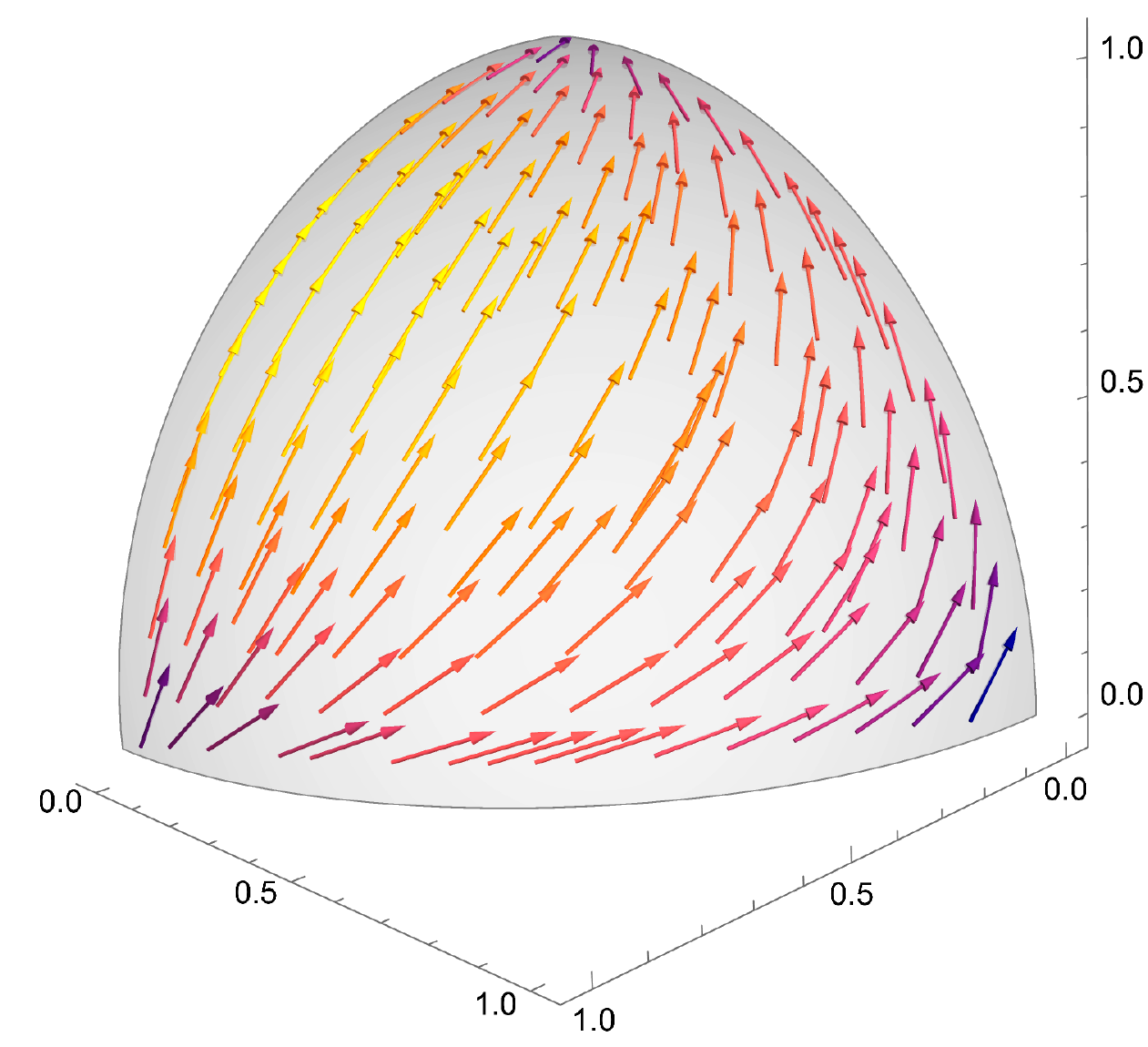}}\hfill     
\caption{\textit{Flow of the negative gradient of the variance}. In the case 
of a decision involving three alternatives, the corresponding state space 
${\mathfrak M}$ is a real projective plane. This two-dimensional manifold 
is not orientable and cannot be embedded in three dimensions, however, 
it can be interpreted as a sphere with antipodal points being identified. 
Thus the flow on ${\mathfrak M}$ can be captured by examining the 
flow on the 
positive octant of the sphere. Here, a three-fold decision is modelled by 
a choice observable ${\hat X}$ with eigenvalues $x_1=1$, $x_2=2$, and 
$x_3=3$. The three axes, corresponding to the three corners of the octant, 
 represent the three eigenstates of ${\hat X}$ 
with no uncertainty. The flow $-\nabla^a V$ generated by the negative 
gradient of the uncertainty (shown in the left panel) takes any initial state 
into one of the states with no uncertainty --- this is the tendency towards 
least surprises. The gradient $\nabla^a X$ of 
the mean is shown on the right panel for comparison. This term is 
multiplied by the Brownian increment $\rd W_t$, which is normally 
distributed with mean zero and variance $\rd t$, so at any moment in 
time the direction of the flow generated by the Brownian fluctuation can 
go either way. 
} 
\label{fig1} 
\end{figure}

Let $\{\psi^a\}$ denote local coordinates for points on ${\mathfrak M}$. A 
point on ${\mathfrak M}$ thus represents a state of mind corresponding 
to a family of vectors $\lambda|\psi\rangle$, $\lambda\neq0$, on Hilbert 
space. For any representative $|\psi\rangle$ of that family corresponding 
to the point $\psi\in{\mathfrak M}$, consider a function on ${\mathfrak M}$ 
through the expectation $X(\psi) = \langle\psi|{\hat X}|\psi\rangle/\langle
\psi|\psi\rangle$. With this convention, and writing $\nabla^a$ for the 
gradient vector, the dynamical equation for the state of mind $|\psi(\xi_t)
\rangle$, when projected down to ${\mathfrak M}$, is given by 
\[ 
\rd \psi^a = - \frac{1}{16}\nabla^a V_X \, \rd t + \frac{1}{4} 
\nabla^a X \, \rd W_t , 
\] 
where $V_X(\psi)= \langle\psi|({\hat X}-X(\psi))^2|\psi\rangle/\langle
\psi|\psi\rangle$ is the function on ${\mathfrak M}$ that corresponds to the 
variance of ${\hat X}$ in the state $|\psi\rangle$. The mathematical 
analysis leading to this result have essentially been provided in 
\cite{Hughston,BH3}, which will not be repeated here. The important 
observation is that the drift term (the coefficient of $\rd t$) that generates 
a tendency flow on the state space ${\mathfrak M}$ is given by the 
negative gradient of the variance (uncertainty). Therefore, on the state 
space there is a tendency of driving the state of mind into one of the 
states with no uncertainty --- this is the flow that attempts to reduce 
surprises. It also follows that if the state of mind is in the vicinity of one 
of the definite states of no uncertainty, then it will be difficult to escape 
from that neighbourhood --- a phenomenon that I referred to as a 
tenacious Bayesian behaviour \cite{Brody2}. I will have more to comment 
on this below. In Figure~\ref{fig1} an example of the negative gradient 
flow of the uncertainty on the state space is sketched, along with the 
gradient flow $\nabla^a X$ of the mean. 

It is worth remarking that in principle properties of the dynamics just 
outlined can be inferred from the squared amplitudes 
\[
\pi_{kt} = \Phi_t^{-1} p_k\, \re^{x_k \xi_t-\frac{1}{2}x_k^2 t} 
\]  
of the coefficients of $|\psi(\xi_t)\rangle$, which evidently contain as much 
information as the state $|\psi(\xi_t)\rangle$ itself. Indeed, starting from the 
expression for $\pi_{kt}$, which can be deduced by use of the Bayes 
formula, one can apply the Ito calculus to deduce the dynamical equation 
satisfied by $\pi_{kt}$, known in communication theory as the Kushner 
equation \cite{Kushner}. However, $\pi_{kt}$ is a martingale, that is, on 
average a conserved process. In particular, the process has no drift (the 
coefficient of $\rd t$ in the Kushner equation for $\pi_{kt}$ is zero), so by 
simply examining the Kushner equation for $\pi_{kt}$ it is difficult to infer key 
properties of the dynamics. In contrast, the surprise-minimising 
feature of the dynamics becomes immediately apparent once the 
process is projected to the state space ${\mathfrak M}$. 

\section*{Difference between psychological and quantum states} 

%\vspace{0.2cm} 
%\noindent \textit{Difference between psychological and quantum states}. 
I have thus far emphasised the similarities in the state of mind as 
represented by a Hilbert space vector (or a density matrix), and the 
physical state of a quantum system as represented by the same scheme. 
There are, however, some important differences. The most important 
one, in my view, can be described through the following example. 
Suppose that the state of a quantum system is very close to one of the 
eigenstates $|x_k\rangle$ of an observable ${\hat X}$, and that the 
measurement outcome yields the value $x_l$, $l\neq k$, for which the 
probability would have been very small. In this case, the interpretation 
of the event is the obvious one: a rare event has occurred. Now 
suppose instead that the state of mind is very close to one of the 
`certain' states $|x_k\rangle$, in a situation where there is a correct 
choice to be made (for example, in deciding what had actually happened 
at an event in the past --- as opposed to choices for which there need 
not be `correct' outputs). In this case, if the correct choice happens to 
be $|x_l\rangle$, $l\neq k$, it does not mean that an unlikely event 
had occurred. Rather, it means that the initial state of mind was a 
misguided one. Putting the matter differently, while a state of a quantum 
system represents the physical reality of the system, a state of mind 
merely represents the person's perception of the state of the world. This 
difference between objective and subjective probabilities has important 
implications discussed below. (There is, of course, the suggestion that the 
state of a quantum system itself is entirely subjective \cite{Fuchs}, but this 
idea will not be explored here.) 

The subjective nature of psychological states gives rise to the following 
challenge. In psychology, it is not uncommon for a group of 
people having varying dispositions be given some information (for example, 
an article to read or a video clip to watch), and their responses are 
examined. One such example is seen in the study of confirmation bias --- a 
bias towards information that confirm their views \cite{Lord,Nickerson}. 
The idea is to investigate 
how people having diverging opinions respond differently to the `same' 
information. The issue here, however, is that the information content of 
the given message such as an article or a video clip is different to people 
with different opinions, even though it is an identical information source. 

To explain this more concretely, consider the simple example considered 
above. Suppose that the preference, or the opinion, of one person on a 
topic is represented by the choice observable ${\hat X}$. Then the 
information-providing observable representing an article discussing this topic 
is given by ${\hat\xi}={\hat X}+{\hat\varepsilon}$. If a second person with a 
different opinion represented by the choice observable ${\hat Y}$ were given 
the same article, then the information-bearing observable for the second 
person is given by ${\hat\eta}={\hat Y}+{\hat\varepsilon}$. They are 
different. Hence, just because two people are given, say, the same article 
to read, to assert that they are given the same information is factually false. 
One important consequence in psychology is that the various conclusions 
drawn from such experiments on how people's behaviour might deviate 
from the rational Bayesian updating require fundamental reexamination. 

The subjective nature of psychological states also gives rise to a 
mathematical challenge. In communication theory, one is typically concerned 
with well-established communication channels, where the signal transmitted 
is assumed to represent an objective reality. Therefore, there is no ambiguity 
in interpreting the information-carrying time series $\{{\hat\xi}_t\}$. However, 
if different receivers were to interpret the `same' message differently, and if 
there is a need to apply statistical analysis on the behaviours of different 
people, then the question arises as to which information process (called 
`filtration' in probability) one should be using for statistical analysis. To my 
knowledge, such a situation has hardly been examined in the vast literature 
of probability and stochastic analysis. 

\section*{Limitation of classical reasoning} 

%\vspace{0.2cm} 
%\noindent \textit{Limitation of classical reasoning}. 
Thus far I have assumed, for definiteness, that all decisions 
are compatible. What this means is that the quantum formalism advocated 
here, while effective, can be reduced, if necessary, to a purely classical 
probabilistic formulation. It seems to me that this assumption does not fully 
reflect the reality, and that it is plausible that not all decisions can be made 
simultaneously by human brains. Indeed, there are empirical examples in 
behavioural psychology that strongly indicate that not all decisions or 
opinions are compatible \cite{Busemeyer,Busemeyer3,Busemeyer4}. 
If so, the observables representing these choices will not commute. 

The issue with the classical updating of likelihoods based on the Bayes 
formula is that it is not well suited to characterise changes of the contexts, 
that is, changes of sample spaces --- represented, for example, by an 
arrival of information that reveals a previously unknown alternative. In 
such a scenario, the prior probability of the new alternative is zero 
(because it was not even known), whereas the posterior can be nonzero. 
Hence, in the language of probability theory, the prior and the posterior 
are not absolutely continuous with respect to each other, prohibiting the 
direct use of the Bayes formula. In contrast, such a change of context can 
be modelled using incompatible observables, along with the von 
Neumann-L\"uders projection postulate. 

To see this, suppose that the prior state of mind is given by $|\psi\rangle = 
\sum_k c_k |x_k\rangle$ when expanded in the eigenstates of ${\hat X}$, 
where $c_m=0$ for some $m$, and suppose that acquisition of information 
takes the form ${\hat\eta} = {\hat Y} + {\hat\varepsilon}$, where ${\hat Y}$ 
cannot be diagonalised using the basis states $\{\!|x_k\rangle\!\}$. 
Then it is possible that the L\"uders state ${\hat\Pi}_\eta|\psi\rangle/\surd{ 
\langle\psi|{\hat\Pi}_\eta|\psi\rangle}$ 
resulting from information acquisition, when expanded in $\{\!|x_k\rangle\!\}$, 
is such that $c_m\neq0$, thus circumventing the constraint of the classical 
Bayes formula. 
Therefore, in a situation whereby choice observables are not compatible, 
the quantum-mechanical formalism proposed here and elsewhere 
\cite{Busemeyer2} becomes a 
necessity, for, the modelling of the dynamical behaviour of a person cannot 
be achieved using the techniques of purely classical probability. 

As a simple example, consider two binary (yes or no) decisions that are 
represented by the choice observables 
\[ 
{\hat X}=\left( \begin{array}{cc} 1 & 0 \\ 0 & -1 \end{array} \right) 
\quad {\rm and} \quad 
{\hat Y}=\left( \begin{array}{cc} \cos\phi & \sin\phi \\ \sin\phi & -\cos\phi 
\end{array} \right) . 
\] 
Evidently, ${\hat X}$ and ${\hat Y}$ cannot be diagonalised simultaneously, 
unless $\phi=0$ (mod $2\pi$). Suppose further that the initial state of mind 
of a person is represented by a Hilbert space vector 
\[ 
|\psi\rangle = \left( \begin{array}{c} \cos\half\theta \\ 
\sin\half\theta \end{array} \right) 
\] 
for some $\theta$. Then the probability that the person giving a `yes' answer 
to question ${\hat X}$ is $\cos^2\frac{1}{2}\theta$; whereas if question 
${\hat Y}$ were asked instead, then the likelihood of giving an affirmative 
answer is $\cos^2\frac{1}{2}(\theta-\phi)$. Note that strictly speaking, 
according to the scheme introduced here the state space for a pair of binary 
decisions is four-dimensional, if the two decisions (questions) are 
simultaneously considered. However, here I am interested in the effect of 
questions being asked sequentially, and for this purpose a two-dimensional 
representation suffices. Thus $|\psi\rangle$ is interpreted to represent an 
abstract state of mind for which a range of binary questions may be asked. 
In quantum theory, such a special state is known as a coherent state. 

Now suppose that question ${\hat Y}$ is asked first, and subsequently 
question ${\hat X}$ is asked. Then from the projection postulate, 
the probability of giving a `yes' answer to question ${\hat X}$, 
irrespective of which answer was given to the first question,  
is $\frac{1}{4}\left(2+\cos(\theta)+\cos(\theta-2\phi)\right)$. For $\phi\neq0$ 
this is different from the \textit{a priori} probability $\cos^2\frac{1}{2}\theta$ 
of answering `yes' to question ${\hat X}$. In other words, the so-called law 
of total probability in the classical probability theory, that the unconditional 
expectation of a conditional expectation equals the unconditional 
expectation, is not applicable when dealing with incompatible propositions. 
Similarly, if 
question ${\hat X}$ is asked before question ${\hat Y}$, then the 
probability of giving a `yes' answer to question ${\hat Y}$ is 
$\cos^2\frac{1}{2}\theta\cos^2\frac{1}{2}\phi + 
\sin^2\frac{1}{2}\theta\sin^2\frac{1}{2}\phi$, which 
is different from  $\cos^2\frac{1}{2}(\theta-\phi)$ when $\phi\neq0$. 

This example is perhaps the simplest one to demonstrate that 
answers to questions can be dependant on the order in which questions 
are asked, provided that the questions are not compatible (see 
\cite{Mayer}, Appendix 2, for a discussion on the order dependence). 
A more elaborate construction of this kind in higher dimension is found in 
\cite{Busemeyer4}. At any 
rate, the violation of the law of total probability shows that this empirical 
phenomenon of order-dependence cannot be explained using compatible 
observables. 

For a pair of binary choices, an attempt is made in \cite{Busemeyer} to 
explain the experiment discussed in \cite{Moore}. The data presented in 
\cite{Moore} show that when people are asked if Clinton is honest, about 
50\% answered `yes', and if they are then asked if Gore is honest, 60\% 
answered `yes'; whereas if the order of the questions are reversed, then 
the figures change into 68\% yes for Gore followed by 60\% yes for Clinton. 
Note however that the explanation of this effect in \cite{Busemeyer} is 
incomplete 
because only conditional probabilities are considered therein, whereas 
the data in \cite{Moore} concern total probabilities. The analysis of total 
probabilities considered here, on the other, shows that by setting $\theta 
\approx 9\pi/20$ and $\phi\approx\pi/15$, the phenomenon reported in 
\cite{Moore} can be explained within the $\pm10\%$ error margin.

\section*{Discussion}

%\vspace{0.2cm} 
%\noindent \textit{Discussion}. 
I have illustrated how the Hilbert-space formalism used in quantum theory is 
highly effective in 
modelling cognitive psychology, in particular, its dynamical aspects. In 
particular, I have shown how an important feature of the dynamics associated 
with Bayesian updating, or equivalently with the von Neumann-L\"uders 
projection, namely, of the uncertainty-reducing trend, is made transparent in 
this formalism. This, in turn, provides an alternative information-theoretic 
perspective on the free 
energy principle, due to the close relation between entropy and variance in 
communication theory. 

One important consequence of the foregoing analysis is that states of low 
uncertainty are always preferred ones, irrespective of whether they represent 
the correct choices. Therefore, if the state of mind happens to be close to one 
of the false choices, then with a rational updating it is difficult to escape from 
this neighbourhood because to achieve this, entropy has to increase before 
it can be decreased again, and this is counter to biological trends 
\cite{Brody1}. In such a situation, it appears that only the accidental effect of 
noise, which otherwise is a nuisance, can rescue the person from the false 
choice within a reasonable timescale, at least when all choices are 
compatible to each other. 

The situation changes once we accept the thesis of \cite{Busemeyer} that 
real-world decisions are never compatible, thus making it a necessity to 
model cognitive behaviours using the quantum formalism. To see this, 
consider a pair of maximally incompatible binary decisions modelled by the 
pair 
\[ 
{\hat X}=\left( \begin{array}{cc} 1 & 0 \\ 0 & -1 \end{array} \right) 
\quad {\rm and} \quad 
{\hat Y}=\left( \begin{array}{cc} 0 & 1 \\ 1 & 0 \end{array} \right) , 
\] 
and suppose that the state of mind is given by 
\[ 
|\psi\rangle = \frac{1}{\sqrt{2}} \left( \begin{array}{c} 1 \\ 
1 \end{array} \right) ,
\] 
or a state very close to $|\psi\rangle$. Because ${\hat Y}|\psi\rangle 
= |\psi\rangle$, this means that the state of mind in relation to decision 
${\hat Y}$ is already fixed to the alternative labelled by the eigenvalue 
$+1$, and that the likelihood of choosing the other alternative labelled by 
the eigenvalue $-1$ is zero, or else very close to zero anyhow. Suppose 
further that the `correct' choice is the one labelled by the eigenvalue $-1$ 
(in a situation where a correct alternative exists). 
The tenacious classical Bayesian behaviour \cite{Brody2} then implies 
that providing partial information ${\hat\eta}={\hat Y}+{\hat\varepsilon}$ 
about the truth will have little impact. Instead, if the person is given 
information, not about the choice ${\hat Y}$, but about ${\hat X}$ in the 
form ${\hat\xi}={\hat X}+{\hat\varepsilon}$, where the magnitude of noise 
${\hat\varepsilon}$ is small, then after acquisition of this information the 
state will change into one of the two possible L\"uders states. These two 
states will be close to one of the two eigenstates of ${\hat X}$. If 
subsequently partial information ${\hat\eta}={\hat Y}+{\hat\varepsilon}$ 
is provided, then irrespective of which L\"uders state is chosen, the state 
of mind will now transform into one that is close to the truth. Thus the 
quantum formalism opens up a new possibility that was unavailable with 
the classical reasonings. 

%
%In this 
%context, if experimental data were to firmly indicate that there are incompatible 
%decisions, then it is of interest to explore whether human behaviours can 
%violate some form of the Bell inequality.  

\vspace{0.5cm}
\begin{footnotesize}
\noindent {\bf Acknowledgements}. The author thanks Bernhard Meister 
for stimulating discussion, and acknowledges support from the EPSRC 
(EP/X019926) and the John Templeton Foundation (grant 62210). The opinions 
expressed in this publication are those of the authors and do not necessarily 
reflect the views of the John Templeton Foundation.
\end{footnotesize}


\vspace{0.1cm}



\begin{thebibliography}{}

\bibitem{Isham} 
Isham,~C. (1995) 
{\em Lectures on Quantum Theory}. 
(London: Imperial College Press). 

\bibitem{Penrose} 
Penrose,~R. (1994) 
{\em Shadows of the Mind}. 
(Oxford: Oxford University Press).

\bibitem{Vitiello} 
Vitiello,~G. (2001) 
{\em My Double Unveiled: The dissipative quantum model of brain}.  
(Amsterdam and Philadelphia: John Benjamin). 

\bibitem{Tegmark} 
Tegmark,~M. (2000) 
Importance of quantum decoherence in brain processes. 
{\em Physical Review} E\textbf{61}, 4194-4206. 

\bibitem{Plotnitsky} 
Plotnitsky,~A. (2014) 
Are quantum-mechanical-like models possible, or necessary, outside 
quantum physics? 
{\em Physica Scripta} \textbf{T163}, 014011. 

\bibitem{Ozawa} 
Ozawa,~M. \& Khrennikov,~A. (2023) 
Nondistributivity of human logic and violation of response replicability 
effect in cognitive psychology. 
{\em Journal of Mathematical Psychology} \textbf{112}, 102739. 

\bibitem{Rao} 
Rao,~C.~R. (1945) 
Information and the accuracy attainable in the estimation of statistical 
parameters. 
{\em Bulletin of Calcutta Mathematical Society} \textbf{37}, 81-91.

\bibitem{Busemeyer} 
Pothos,~E.~M. \& Busemeyer,~J.~R. (2013) 
Can quantum probability provide a new direction for cognitive modeling? 
{\em Behavioral and Brain Sciences} \textbf{36}, 255-327. 

\bibitem{Busemeyer2} 
Pothos,~E.~M. \& Busemeyer,~J.~R. (2022) 
Quantum cognition.  
{\em Annual Reviews of Psychology} \textbf{73}, 749-778. 

\bibitem{Busemeyer0} 
Busemeyer,~J.~R. \& Bruza,~P.~D. (2012) 
{\em Quantum Models of Cognition and Decision}. 
(Cambridge: Cambridge University Press). 

\bibitem{Khrennikov} 
Haven,~E. \& Khrennikov,~A. (2016) 
Statistical and subjective interpretations of probability in quantum-like 
models of cognition and decision making. 
{\em Journal of Mathematical Psychology} \textbf{74}, 82-91. 

\bibitem{Wiener} 
Wiener,~N. (1948) 
{\em Cybernetics, or Control and Communication in the Animal and the 
Machine}.  
(Boston: The Technology Press of the MIT).

\bibitem{Friston0} 
Friston,~K., Kilner,~J. \& Harrison,~L. (2006) 
A free energy principle for the brain. 
{\em Journal of Physiology -- Pairs} \textbf{100}, 70-87. 

\bibitem{Brody2} 
Brody,~D.~C. (2022) 
Noise, fake news, and tenacious Bayesians. 
{\em Frontiers in Psychology} \textbf{13}, 797904.

\bibitem{DeGroot} 
DeGroot,~M.~H. (1970) 
{\em Optimal Statistical Decisions} 
(New York: McGraw-Hill).

\bibitem{Berger} 
Berger,~J.~O. (1985) 
{\em Statistical Decision Theory and Bayesian Analysis} 
(New York: Spinger-Verlag).

\bibitem{BH} 
Brody,~D.~C. \& Hook,~D.~W. (2009) 
Information geometry in vapour-liquid equilibrium. 
{\em Journal of Physics} A\textbf{42}, 023001. 

\bibitem{KF1} 
Friston,~K. (2010) 
The free-energy principle: a unified brain theory? 
{\em Nature Reviews Neuroscience} \textbf{11}, 127-138. 

\bibitem{BH2} 
Brody,~D.~C. and Hughston,~L.~P. (2006) 
Quantum noise and stochastic reduction. 
{\em Journal of Physics} A\textbf{39}, 833-876.

\bibitem{Kailath} 
Kailath,~T. (1968) 
An innovations approach to least-squares estimation. Part I: Linear 
filtering in additive white noise.  
{\em IEEE Transactions on Automatic Control} \textbf{13}, 646-655.

\bibitem{Wonham}
Wonham,~W.~M. (1965) 
Some applications of stochastic differential equations to optimal nonlinear 
filtering. 
{\em Journal of the Society for Industrial and Applied Mathematics: Control} 
A\textbf{2}, 347-369. 
%(doi:10.1137/0302028)

\bibitem{KF2} 
Fields,~C., Friston,~K., Glazebrook,~J.~F. \& Levin,~M.  (2022) 
A free energy principle for generic quantum systems. 
{\em Progress in Biophysics and Molecular Biology} \textbf{173}, 36-59. 

\bibitem{BH4} 
Brody,~D.~C. and Hughston,~L.~P. (1999) 
Geometrisation of statistical mechanics. 
{\em Proceedings of the Royal Society London} A\textbf{455}, 1683-1715.

\bibitem{Hughston} 
Hughston,~L.~P. (1996) 
Geometry of stochastic state vector reduction. 
{\em Proceedings of the Royal Society London} A\textbf{452}, 953-979.

\bibitem{BH3} 
Brody,~D.~C. and Hughston,~L.~P. (2002) 
Stochastic reduction in nonlinear quantum mechanics. 
{\em Proceedings of the Royal Society London} A\textbf{458}, 1117-1127.

\bibitem{Kushner} 
Kushner,~H.~J. 
On the differential equations satisfied by conditional probability densities of 
Markov processes, with applications. 
{\em Journal of the Society for Industrial and Applied Mathematics: Control} 
A\textbf{2}, 106-119 (1964). 

\bibitem{Fuchs} 
Fuchs,~C.~A. \& Schack,~R. (2014) 
QBism and the Greeks: why a quantum state does not represent an 
element of physical reality. 
{\em Physica Scripta} \textbf{90}, 015104. 

\bibitem{Lord} 
Lord,~C.~G., Ross,~L. \& Lepper,~M.~R. (1979) 
Biased assimilation and attitude polarization: The effects of prior theories on 
subsequently considered evidence. 
{\em Journal of Personality and Social Psychology} \textbf{37}, 2098-2109. 

\bibitem{Nickerson} 
Nickerson,~R.~S. (1998) 
Confirmation bias: A ubiquitous phenomenon in many guises. 
{\em Review of General Psychology} \textbf{2}, 175-220.

\bibitem{Busemeyer3} 
Pothos,~E.~M. \& Busemeyer,~J.~R. (2009) 
A quantum probability explanation for violations of `rational' decision theory.  
{\em Proceedings of the Royal Society} B\textbf{276}, 2171-2178. 

\bibitem{Busemeyer4} 
Basieva,~I., Pothos,~E.~M., Trueblood,~J., Khrennikov,~A. \& 
Busemeyer,~J. (2017) 
Quantum probability updating from zero priors (by-passing Cromwell’s rule). 
{\em Journal of Mathematical Psychology} \textbf{77}, 58-69. 

\bibitem{Mayer} 
Mayer,~P.~A. (1995) 
{\em Quantum Probability for Probabilists}. 2nd ed. 
(Berlin: Springer). 

\bibitem{Moore} 
Moore,~D.~W. (2002) 
Measuring new types of question-order effects: Additive and subtractive. 
{\em The Public Opinion Quarterly} \textbf{66}, 80-91. 

\bibitem{Brody1} 
Brody,~D.~C. \& Trewavas,~A.~J. (2022) 
Biological efficiency in processing information. 
arXiv:2209.11054.

\end{thebibliography}
\end{document}